\documentclass[a4paper,twocolumn,11pt]{quantumarticle}
\pdfoutput=1
\usepackage[utf8]{inputenc}
\usepackage[english]{babel}
\usepackage[T1]{fontenc}
\usepackage{amsmath}
\usepackage{hyperref}
\usepackage{graphicx}\usepackage{braket}
\usepackage{amsmath,amsfonts,amssymb,amstext,amscd,amsthm}
\usepackage{tikz}
\usepackage{lipsum}

\begin{document}

\title{Quantum interference and coherence in one-dimensional disordered and localized quantum walk}

\author{Shivani Singh}
\email{shivanis@imsc.res.in}
\affiliation{The Institute of Mathematical Sciences, C. I. T campus, Taramani, Chennai, 600113, India}
\affiliation{Homi Bhabha National Institute, Training School Complex, Anushakti Nagar, Mumbai 400094, India}
\author{C. M. Chandrashekar}
\email{chandru@imsc.res.in}
\affiliation{The Institute of Mathematical Sciences, C. I. T campus, Taramani, Chennai, 600113, India}
\affiliation{Homi Bhabha National Institute, Training School Complex, Anushakti Nagar, Mumbai 400094, India}

\maketitle
\begin{abstract}

One-dimensional discrete-time quantum walk has played an important role in development of quantum algorithms and protocols for different quantum simulations. The speedup observed in quantum walk algorithms is attributed to quantum interference and coherence of the wave packet in position space. Similarly, localization in quantum walk due to disorder is also attributed to quantum interference effect. Therefore, it is intriguing to have a closer look and understand the way quantum interference manifests in different forms of quantum walk dynamics. Quantum coherence in the system is responsible for quantum interference in the system. Here we will use coherence measure to quantify the interference in the discrete-time quantum walk. We show coherence in the position and coin space, together and independently, and present the contribution of coherence to the quantum interference in the system. This study helps us to differentiate the localization seen in one dimensional discrete-time quantum walks due to different forms of disorders and topological effects. 

\end{abstract}


\maketitle

\section{Introduction} 
Interference of quantum states, traversing multiple paths in parallel, has played an important role in quantum cryptography\,\cite{7, 8}, quantum metrology\,\cite{9}, interferometry\,\cite{10, 11}, and various other quantum information processing tasks\,\cite{12, 13}.  Though interference is widely studied and a universal theory of it is know\,\cite{6}, the intricacy involved in the dynamics during interference of quantum state and the way it can be quantified is still a topic of interest\,\cite{31,32}. 

Quantum walk (QW), developed using the aspects of quantum mechanics, spreads quadratically faster in position space when compared to its classical counterpart, classical random walk\,\cite{Ria1958, Fey86, ADZ93, Mey96, Kem03, Ven12}. QW has been used to model dynamics in many system such as photosynthesis\,\cite{14, 15}, breakdown of electric field driven system\,\cite{16, 17}, diffusion in quantum system\,\cite{18}, and localization\,\cite{Joy12, AVM11}. Universal computation\,\cite{Chi09,LCE10,SCASC20,SCAC19} and quantum simulations\,\cite{CBS10,Cha13,MBD13,MMC17} are some of the other important directions where QW is considered to be one of the powerful quantum algorithmic tool to establish quantum supremacy.  

In QW, superposition, coherence and interference plays an important role in observing speedup in dynamics and in generation of entanglement\,\cite{3, 4}. It is also one of the system where dynamics can be controlled to modify the way interference manifests leading to different interesting phenomena. That is, one can realize a ballistic spread of wave packet in position space and at the same time realize strong localization and weak localization by modifying the dynamics\,\cite{Joy12,21}. Therefore, quantifying and understanding interference in QW will play an important role in exploring the possibility of explicitly using quantum interference with the help of coherence measure in quantum information processing tasks. Quantum interference between different computational paths, which plays an important role in quantum computation, will also get a better understanding which could result in interest towards optimization of computational tasks. From the theory of  Anderson localization\,\cite{22} and weak localization\,\cite{23} , we know that broken symmetry in the dynamics of system due to disordered media leads to localization of energy states and it has been experimentally verified\,\cite{24}.  It is also established that the quantum interference is what results in Anderson localization and this effect is absent in classical systems.  

In one-dimensional discrete-time QW, quantum interference leads to both, ballistic spread in the position space of homogeneous evolution and localization in disordered evolution. One-dimensional discrete-time QW is defined on a composite Hilbert space of coin and position. Similarly, topological phases has also been engineered in discrete-time QW as presented in pervious publications\,\cite{25,26,27,28} with topological localization. Topological phases are very important in experimental realization of topological insulators. These phases do not break any symmetry, rather are described by the presence and absence of certain symmetries such as time-reversal, particle-hole and chiral symmetry\,\cite{26}. In one dimension, the strongest localization is obtained in topological case which has also been realised on one-dimensional discrete-time QW\,\cite{28}.  

In this work, we have quantified the interference in terms of coherence in the system. We have shown that coherence in the coin space quantifies the wave nature of the walker and coherence in the position space quantifies the particle nature of the walker, respectively in one dimensional discrete-time QW. Coherence in the coin-space also shows the signature of alternate constructive and destructive interference in localised case. Section \ref{Sec2}, discuss discrete-time QW and its localization in one-dimension. Section \ref{Sec3} quantifies coherences and then interference in terms of coherence in QW by quantifying wave and particle nature, respectively. Section \ref{Sec4} discuss in detail the interference in homogeneous QW and localized QW and Section \ref{Sec5} gives the concluding remark.  


\section{\label{Sec2}Discrete-time QW and localization}

 {\textbf{Homogeneous discrete-time QW:}} Discrete-time QW in one-dimension is defined on composite Hilbert space $\mathcal{H} = \mathcal{H}_c \otimes \mathcal{H}_p$. $\mathcal{H}_c$ is the coin Hilbert space with two internal degree of freedom, $span\{\ket{\uparrow},\ket{\downarrow}\} $ as the basis states and $\mathcal{H}_p$ is the position Hilbert space defined by the basis states $\ket{x}$ where $x \in \mathbb{I}$. Each step of discrete-time QW is evolved using a quantum coin operation,
\begin{equation}
B(\theta) \otimes \mathbb{I} \equiv \begin{pmatrix}
\cos\theta & \sin\theta\\
\sin\theta & -\cos\theta
\end{pmatrix} \otimes \sum_x \ket{x}\bra{x}
\end{equation}
followed by a  position shift operation
\begin{equation}
S_x \equiv \sum_x \left[\ket{\uparrow}\bra{\uparrow} \otimes \ket{x-1}\bra{x} + \ket{\downarrow}\bra{\downarrow} \otimes \ket{x+1}\bra{x}\right],
\end{equation}
which evolves the particle into superposition of its basis states. Therefore, the unitary evolution operation at each step is given by $
W_x(\theta) \equiv S_x \left[B(\theta) \otimes \mathbb{I}\right]$
and after $t$-time steps, state of the system is $\ket{\Psi_t} = W_x(\theta)^t \ket{\Psi_{in}}$ for a given initial state $\ket{\Psi_{in}} = (\alpha |\uparrow\rangle + \beta |\downarrow\rangle)\otimes |x=0\rangle$. The coin parameter $\theta$ controls the variance of the probability distribution in position space. Fig. \ref{fig2} shows the probability distribution for homogeneous discrete-time QW (HQW) when $\theta = \pi/4$ after 100 steps of walk.


{\textbf{Disorder induced localization:}} Disorder is introduced in the discrete-time QW evolution by using randomized quantum coin operation. Anderson localization can be simulated using spatial disorder in QW and  weak localization can be simulated by using temporal disorder in QW. 


{\textit{Spatial Disorder:}} The spatial disorder in QW (SQW) evolution is introduced by a position dependent coin operation $B(\theta_x)$ in the evolution dynamics, where parameter $\theta_x$ is randomly picked for each position from the range $0 \leq \theta_x \leq \pi$. The state after time  $t$ with spatial disorder in single parameter evolution operation will be, 
\begin{align}
\ket{\Psi_t}_S &= \left[W_x(\theta_{x})\right]^t \ket{\Psi_{in}} \nonumber\\
&= \left[S_x \left(\sum_{x} B(\theta_x) \otimes \ket{x}\bra{x}\right)\right]^t \ket{\Psi_{in}},
\end{align}
The iterative form of the state of the walker at each position $x$ and at time $t+1$ with spatial disorder will be,
\begin{align}\label{eq5}
\begin{pmatrix}
\psi^{\uparrow}_{x,t+1}\\
\psi^{\downarrow}_{x,t+1}
\end{pmatrix} &= 
\begin{pmatrix}
\cos\theta_{x+1} & \sin\theta_{x+1} \\
0 & 0
\end{pmatrix} 
\begin{pmatrix}
\psi^{\uparrow}_{x+1,t}\\
\psi^{\downarrow}_{x+1,t}
\end{pmatrix}  \nonumber \\ 
& + \begin{pmatrix}
0 & 0 \\
\sin\theta_{x-1} & -\cos\theta_{x-1}
\end{pmatrix} 
\begin{pmatrix}
\psi^{\uparrow}_{x-1,t}\\
\psi^{\downarrow}_{x-1,t}
\end{pmatrix}.
\end{align}
In general, spatial disorder induces a strong localization of the walker in position space with time in QW evolution\,\cite{21}. This can also be seen in Fig.\,\ref{fig2}.

        
{\textit{Temporal Disorder:}} The temporal disorder in QW (TQW) evolution is introduced by a time dependent coin operation $B(\theta_t)$ in the evolution dynamics, where parameter $\theta_t$ is randomly picked for each time-step from the range $0 \leq \theta_t \leq \pi$. The state in temporal disordered system after time $t$, using a single parameter evolution operation will be,
\begin{align}
\ket{\Psi_t}_T = W_x(\theta_t)...W_x(\theta_2)W_x(\theta_1)\ket{\Psi_{in}}.
\end{align} 
The iterative form of the state of the walker at each position $x$ and time $(t+1)$ will be identical to Eq.\,\eqref{eq5} with a replacement of $\theta_t$ in place of $\theta_{x\pm 1}$. Temporal disorder induces a weak localization in position space of QW evolution\,\cite{33, 34} and Fig.\,\ref{fig2} shows localization due to temporal disorder.


{\textbf{Topological localization:}} Topological phases are associated with the presence of symmetries such as time-reversal symmetry, particle-hole symmetry and chiral symmetry. In one dimensional discrete-time QW, all the three symmetries has been achieved and well studied using the split-step QW form and therefore topological phases has also been engineered\,\cite{28}. Topological property can be introduced by considering a QW with each step split into two with different coin parameters $\theta_i$ as
\begin{equation}
W(\theta_1, \theta_2) = S_+R_{\theta_2}S_-R_{\theta_1},
\end{equation}
where,
\begin{align}
R_{\theta} &\equiv B(\theta) \otimes \mathbb{I}, ~~~~ 0<\theta \leq \pi
\end{align}
and for which the position split shift operators are,
\begin{align}         
S_- &= \ket{\uparrow}\bra{\uparrow} \otimes \ket{x-1}\bra{x} + \ket{\downarrow}\bra{\downarrow} \otimes \ket{x}\bra{x};\\
S_+ &= \ket{\uparrow}\bra{\uparrow} \otimes \ket{x}\bra{x} + \ket{\downarrow}\bra{\downarrow} \otimes \ket{x+1}\bra{x}.
\end{align}               
To create a real space boundary between topologically distinct phases and reveal non-trivial topological properties at the interface, one can choose different $\theta_2$ to the left $(R_{\theta_{2-}})$ and right side $(R_{\theta_{2+}})$ of the intial state (interface) in the position space, while defining the coin operation $R_{\theta_1}$ uniformly on the entire position space. Different combinations of $\theta_1 $ and $\theta_2$ gives different probability distribution across the interface but maximum localization at the interface can be achieved for the combination $(\theta_1,\theta_{2-},\theta_{2+}) = (-3\pi/4,-\pi/2,\pi/2)$ as seen in Fig.\,\ref{fig7}-(g). Probability distribution for different combinations of $(\theta_1, \theta_{2-}, \theta_{2+})$ is shown in Fig.\,\ref{fig7}. For parameters when topological edges are created (different winding number) we see a localized state, Fig.\,\ref{fig7}-(c) and (g)\,\cite{28}.
%
%
\begin{figure}
	\centering
	\includegraphics[width=0.9\columnwidth]{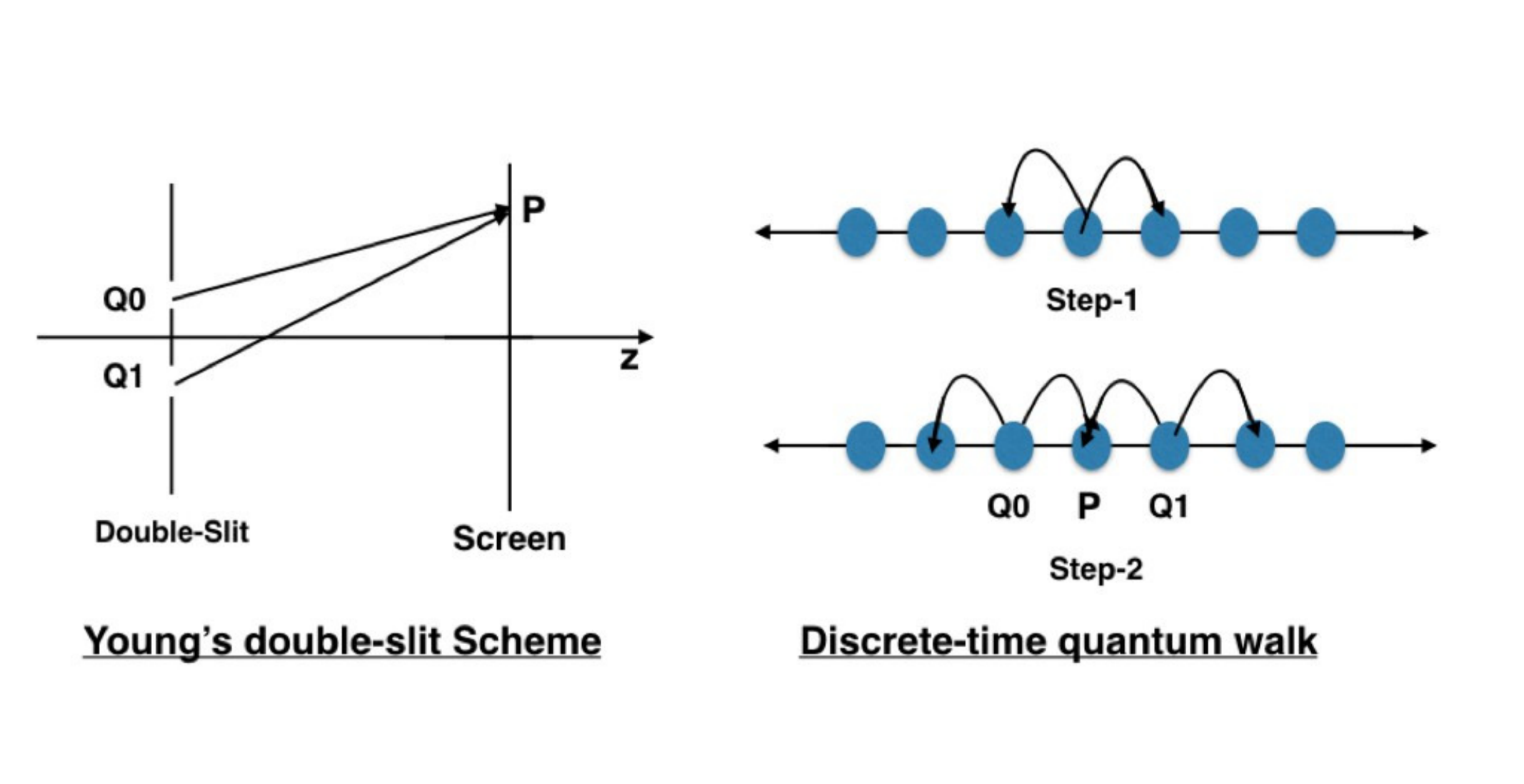} 
\caption{A comparison of double-slit experiment with QW. In double-slit scheme, a photon impinges on the slits and renders into two possible paths which are detected on the screen. In QW, in the second step the probability of finding a particle at the point P is an interference of probability amplitude from Q0 and Q1. Therefore, QW as a whole can be seen as a multi-slit experiment with increase in number of slits with time.}
	\label{fig1}
\end{figure}
\begin{figure}
	\centering		
     \includegraphics[width=0.49\columnwidth]{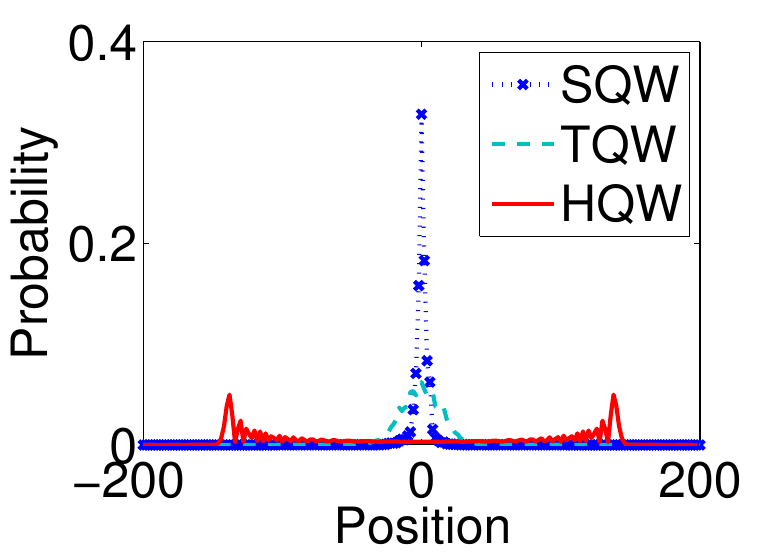}
     \includegraphics[width=0.49\columnwidth]{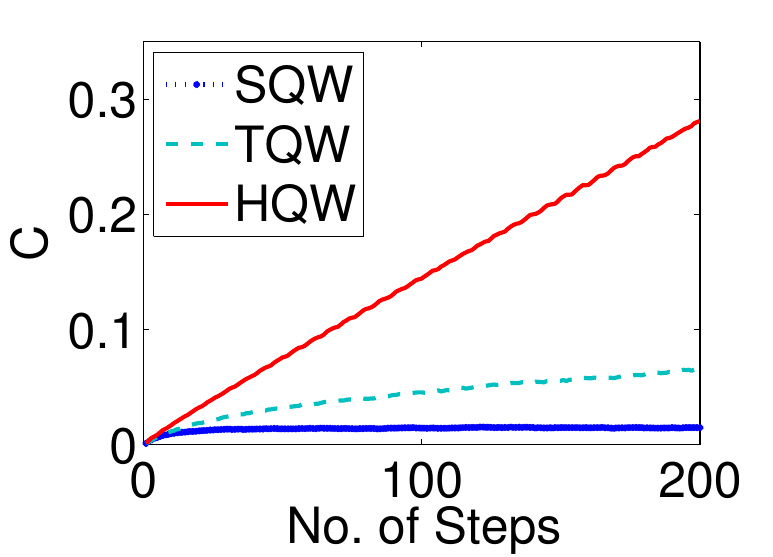}  
	\caption{left to right, first plots shows the probability distribution for one-dimensional homogeneous QW (HQW), spatial-disordered QW (SQW) and temporal-disordered QW (TQW) after 200 time-steps, respectively. Second plot shows coherence measure ($C(\rho) \equiv C$) in complete Hilbert space $\mathcal{H} = \mathcal{H}_c \otimes \mathcal{H}_p$ w.r.t. time-steps for HQW, SQW and TQW. The initial state of the particle is $\dfrac{1}{\sqrt{2}}(\ket{\uparrow} + \textit{i}\ket{\downarrow}) \otimes \ket{x=0}$ and disordered system is averaged over 100-runs. A steep increase in coherence is seen for HQW and saturation at a very small value for SQW with time. For  TQW, slow but continuous increase of coherence is seen with time.}
	\label{fig2}
\end{figure}
\begin{figure}
	\centering
	\includegraphics[width=0.49\columnwidth]{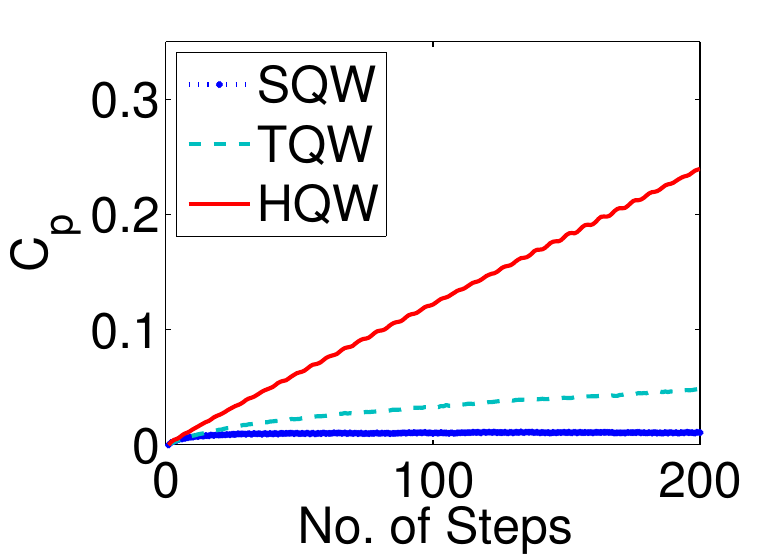}
    \includegraphics[width=0.49\columnwidth]{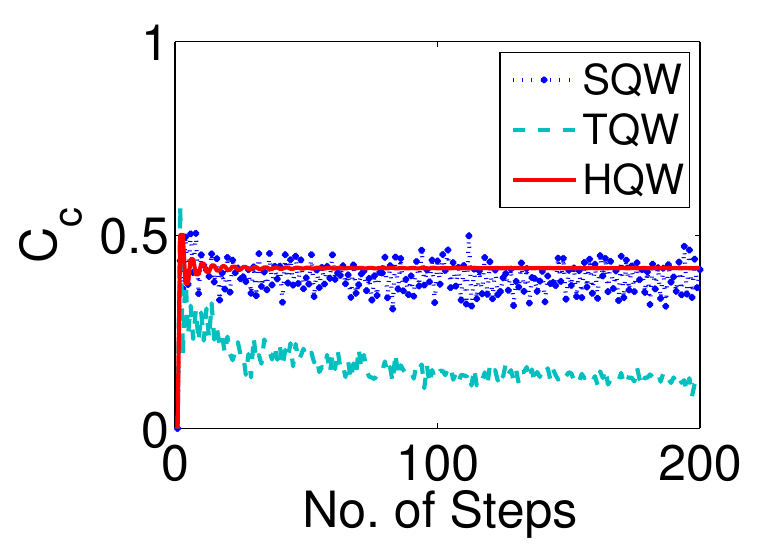}
	\caption{Left to right, first plot shows the coherence measure in position Hilbert space ($C(\rho_p) \equiv C_p$) and second plot show coherence measure in coin Hilbert space ($C(\rho_c) \equiv C_c$) with time for HQW, SQW and TQW. The initial state of the walker is $\dfrac{1}{\sqrt{2}}(\ket{\uparrow} + \textit{i}\ket{\downarrow}) \otimes \ket{x=0}$ and disordered system is averaged over 100-runs.}
\label{fig3}	
\end{figure}
\begin{figure}[h!]
	\centering
	\includegraphics[width=0.9\linewidth]{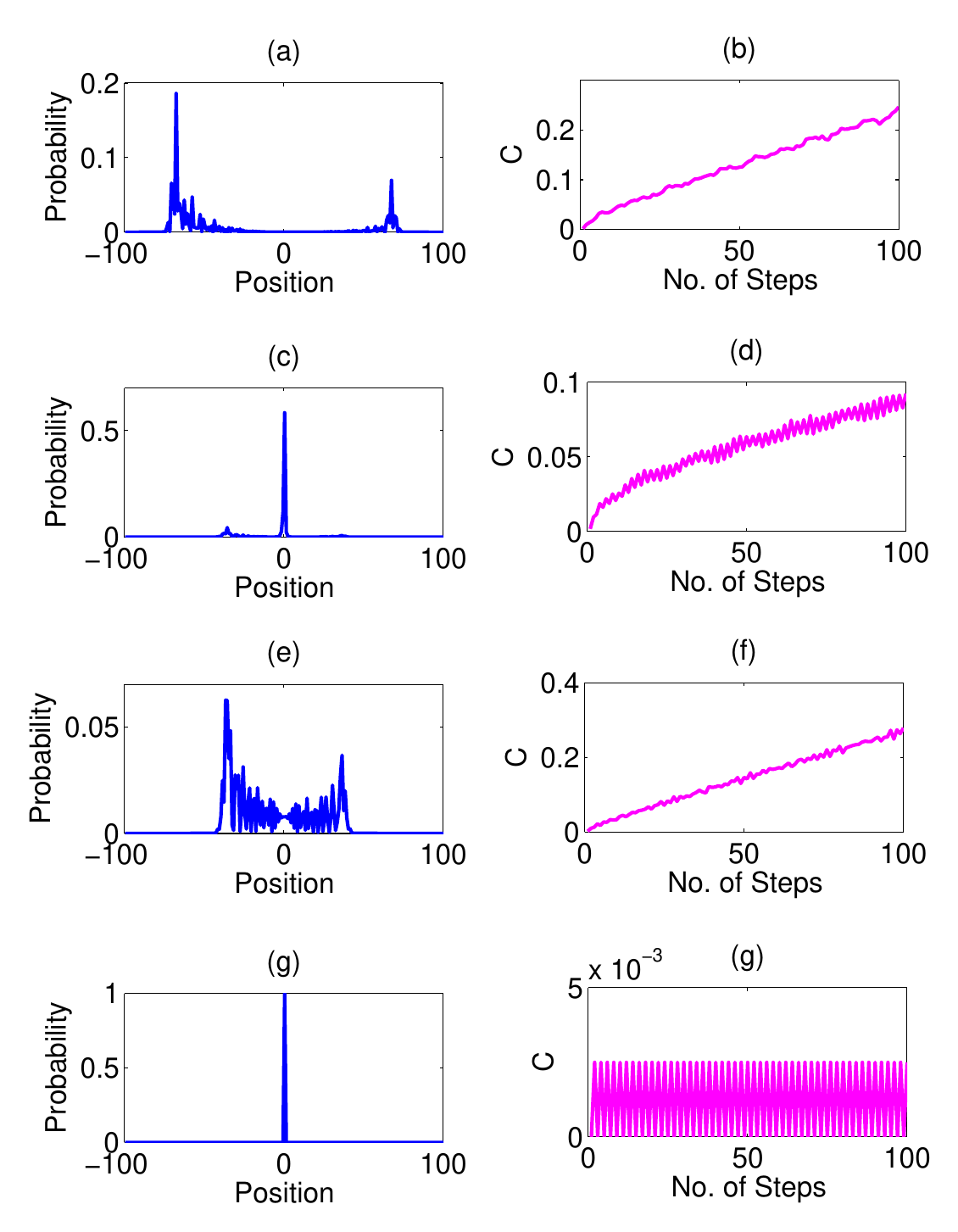} 
	\caption{(a), (c), (e) and (g) represents the probability distributions of the split-step QW and (b), (d), (f) and (h) represents the coherence measure ($C(\rho) \equiv C$) in Hilbert space $\mathcal{H} = \mathcal{H}_c \otimes \mathcal{H}_p$, after 100 step. Initial state is $\ket{\psi_{in}} = \dfrac{1}{\sqrt{2}}(\ket{\uparrow}+ \ket{\downarrow}) \otimes \ket{\textit{x} = 0}$. For (a) and (b) $\theta_1 = \pi/4$ and ($\theta_{2-}, \theta_{2+}$) = ($-\pi/8, \pi/8$), (c) and (d) $\theta_1 = \pi/4$ and ($\theta_{2-}, \theta_{2+}$) = ($-3\pi/8, 3\pi/8$), (e) and (f) $\theta_1 = -3\pi/4$ and ($\theta_{2-}, \theta_{2+}$) = ($5\pi/8, 3\pi/8$) and for (g) and (h) $\theta_1 = -3\pi/4$ and ($\theta_{2-}, \theta_{2+}$) = ($-\pi/2, \pi/2$).}
	\label{fig7}
\end{figure}
\begin{figure}
	 \centering
	\includegraphics[width=0.9\linewidth]{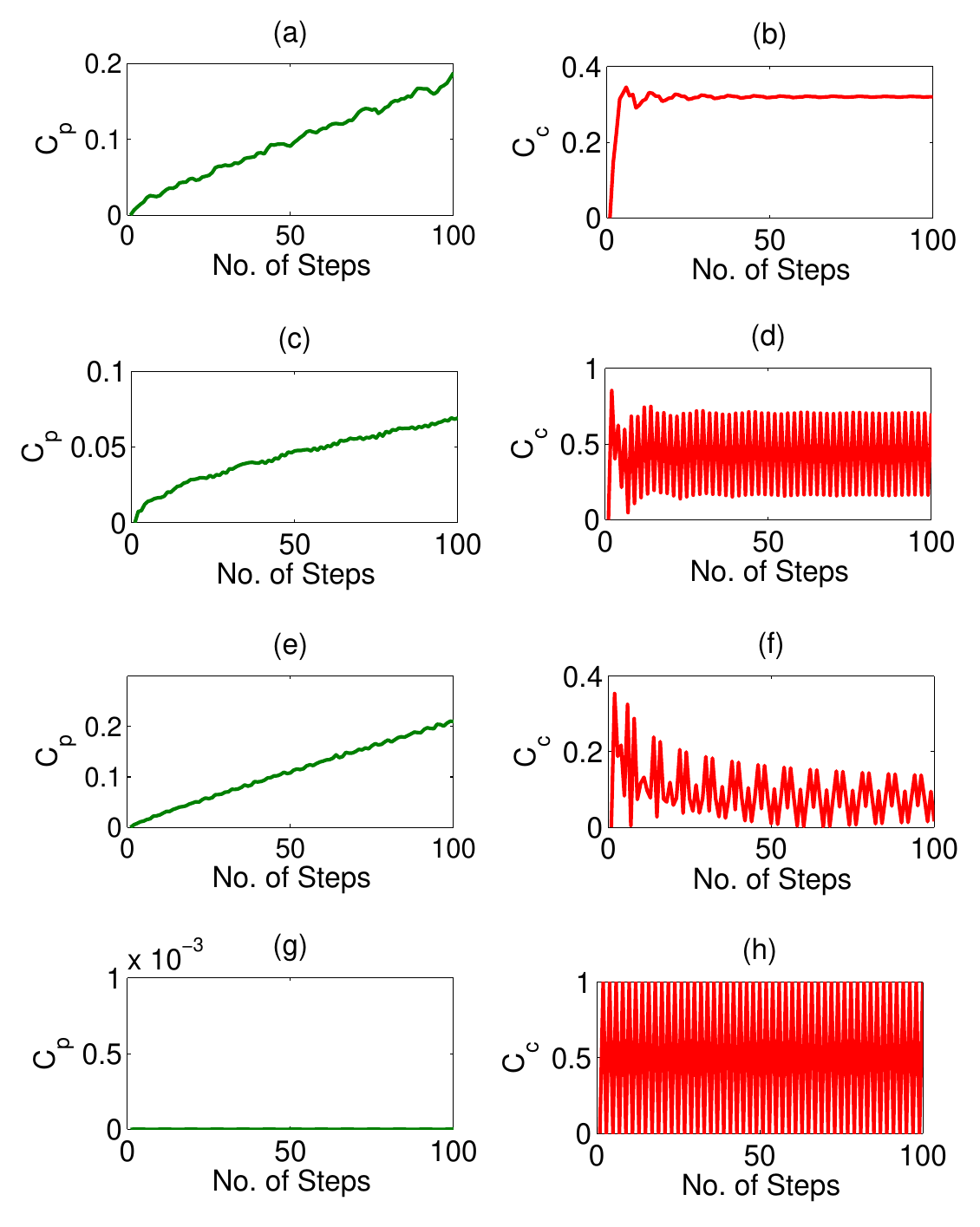} 
	\caption{(a), (c), (e) and (g) represents the coherence measure in position Hilbert space ($C(\rho_p) \equiv C_p$) of the split-step QW and (b), (d), (f) and (h) represents the coherence measure in coin Hilbert space ($C(\rho_c) \equiv C_c$) after 100 time-steps with initial state $\ket{\psi_{in}} = \dfrac{1}{\sqrt{2}}(\ket{\uparrow}+ \ket{\downarrow}) \otimes \ket{\textit{x} = 0}$, respectively. For (a) and (b) $\theta_1 = \pi/4$ and ($\theta_{2-}, \theta_{2+}$) = ($-\pi/8, \pi/8$), (c) and (d) $\theta_1 = \pi/4$ and ($\theta_{2-}, \theta_{2+}$) = ($-3\pi/8, 3\pi/8$), (e) and (f) $\theta_1 = -3\pi/4$ and ($\theta_{2-}, \theta_{2+}$) = ($5\pi/8, 3\pi/8$) and, (g) and (h) $\theta_1 = -3\pi/4$ and ($\theta_{2-}, \theta_{2+}$) = ($-\pi/2, \pi/2$).}
	\label{fig6}
\end{figure}
%

\section{\label{Sec3} Quantum coherence and interference in QW}

In QW, probability amplitudes coherently superimpose while propagating and generates interference at different sites. Quantum interference is to explore the amount of wave or particle nature at the same time. According to complementarity relation for interference, the knowledge of which-path information limits the interference visibility in an interference experiment\,\cite{GY88, E96}. However, visibility implies lack of particle nature and which-path information implies lack of wave nature. Quantum coherence can be used as to quantify the wave nature and which-path information in the discrete-time quantum walk. 

Classically, coherence of an optical field can be seen as the ability to produce interference as shown in schematic view of the $N$-path interference when $N=2$ in Fig.\,\ref{fig1}. Degree of spatial coherence of light for two-path intereference experiment in context of quantum theory has already been derived\,\cite{30}.  
In quantum mechanics, all the diagonal density matrices in the  Hilbert space of basis $\ket{i}$ such that $i \in \mathbb{I}$ are incoherent\,\cite{32}. Therefore, coherence for a given density matrix can be defined by the off-diagonal elements of the matrix. 
The normalised coherence measure for a density matrix $\rho$ with dimensionality $N$ of the Hilbert space is given by,
\begin{equation}\label{eq14}
C \equiv C(\rho) = \dfrac{1}{N-1}\sum_{i \neq j} \left| \bra{i}\rho\ket{j} \right| = \dfrac{1}{N-1}\sum_{i \neq j} \left| \rho_{ij} \right|.
\end{equation}
In discrete-time QW, if dimensionality of the position Hilbert space is $N$ and the dimensionality of the coin Hilbert space is $2$, then the dimensionality of the complete Hilbert space of QW is $2N$ in the Eq.\,\eqref{eq14}. The normalised coherence measure above is same as the coherence measure based on the quantifier of quantum coherence\,\cite{31, 32}. 

Quantum coherence can be used to quantify the interference in the system\,\cite{MQSP15}. For QW set-up, the internal state of walker forms the incoherent bases at each position. Therefore, wave nature can be quantified in terms of quantum coherence of the internal state of the walker (or coin Hilbert state) given by,
\begin{align}
 C(\rho_c) &= \sum_{i \neq j} \left| \rho_{c_{ij}} \right| \nonumber \\
&= \sum_{i \neq j} \left| (Tr_p(\rho))_{ij} \right| \nonumber \\
&\equiv C_c .
\end{align}
 and $\dfrac{1}{N}$ times of $C(\rho_c)$ will quantify the wave nature of the walker over the whole position space of size $N$ after $t \approx N/2$-time steps. The particle nature is given by which-way information and can be quantified by the upper bound of the success probability in the position state of the quantum walk which is given by,
\begin{align} \label{particleNature}
D_{Q} &= 1- \dfrac{1}{N-1}\sum_{i \neq j} \left| \rho_{p_{ij}} \right| \nonumber \\
&= 1 - C(\rho_p) \nonumber \\
&= 1 - C_p
\end{align}
and the duality relation for the system is given by $\Big(D_Q + \dfrac{1}{N} C(\rho_c)\Big) \leq 1$.

An isolated one step view of discrete-time QW at position $x$ due to nearest neighbouring site, in one dimension, has a similar set-up as the Young's double-slit experiment as shown in Fig.\,\ref{fig1}. For double-slit experiment, interference visibility is defined as,
\begin{align}
\mathcal{V} &\equiv \frac{I_{max} - I_{min}}{I_{max} + I_{min}}
\end{align}
 where, $I_{max}$ and $I_{min}$ is the maximum and minimum intensity in the neighbouring fringes. In discrete-time QW, the intensity is replaced by the probability of finding the walker at some point $x$ in position space at a given time $(t+1)$. 

Probability to find walker at position $x$ at time $(t+1)$ is due to the superposition of the probability amplitudes at the nearest neighbours at time $t$. The components of probability amplitude to find walker at position $x$ at time $(t+1)$ is given by,
\begin{align}
\begin{pmatrix}
\psi_{x,(t+1)}^{\uparrow} \\
\psi_{x,(t+1)}^{\downarrow}
\end{pmatrix} &= \begin{pmatrix}
\cos\theta_{x+1} & \sin\theta_{x+1} \\
0 & 0
\end{pmatrix} 
\begin{pmatrix}
\psi^{\uparrow}_{x+1,t}\\
\psi^{\downarrow}_{x+1,t}
\end{pmatrix}  \nonumber \\ 
& + \begin{pmatrix}
0 & 0 \\
\sin\theta_{x-1} & -\cos\theta_{x-1}
\end{pmatrix} 
\begin{pmatrix}
\psi^{\uparrow}_{x-1,t}\\
\psi^{\downarrow}_{x-1,t}
\end{pmatrix}.
\end{align}

\begin{figure}
	\centering
	\includegraphics[width=0.95\columnwidth]{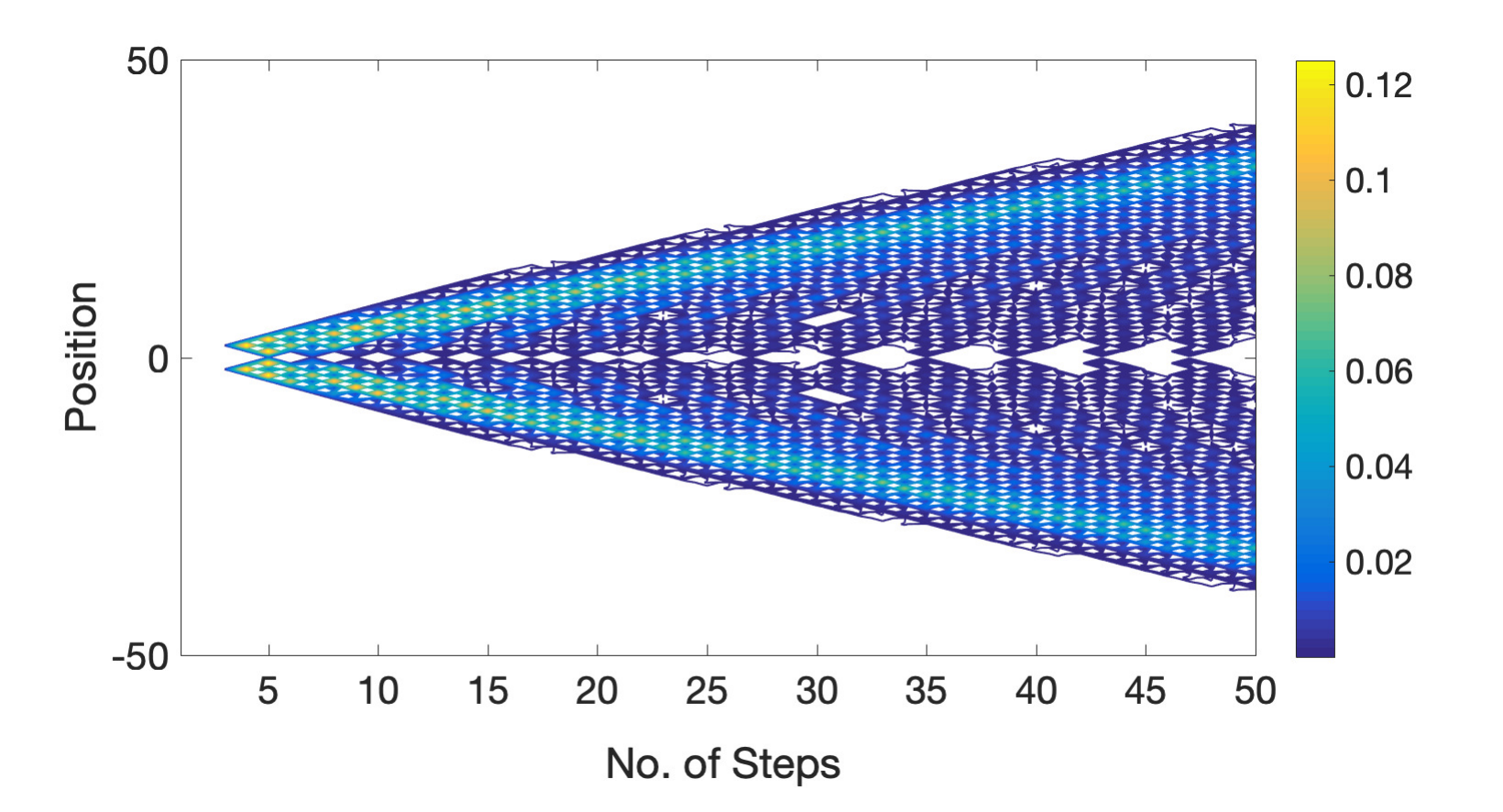}
	\includegraphics[width=0.95\columnwidth]{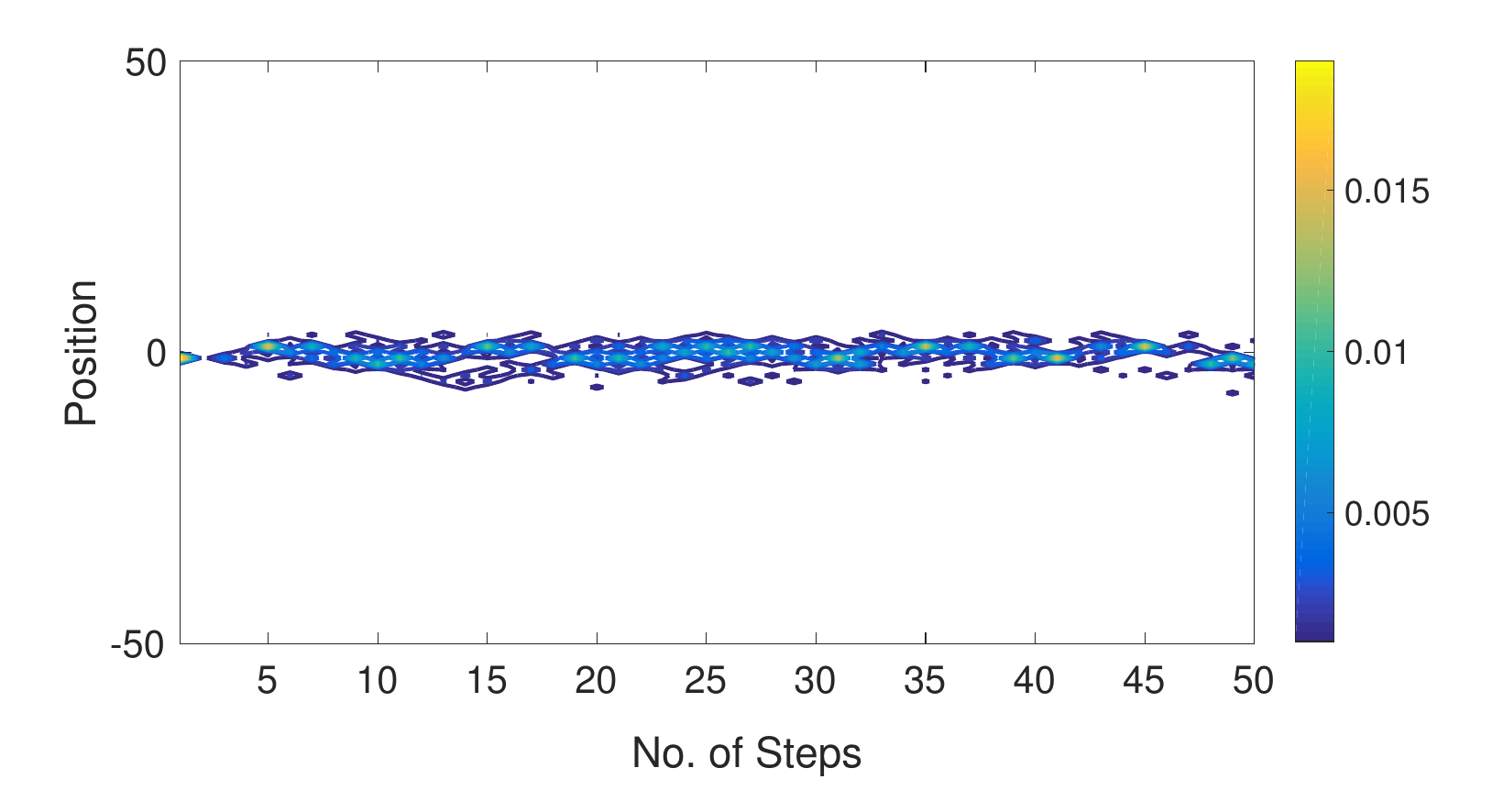} 
 	\includegraphics[width=0.95\columnwidth]{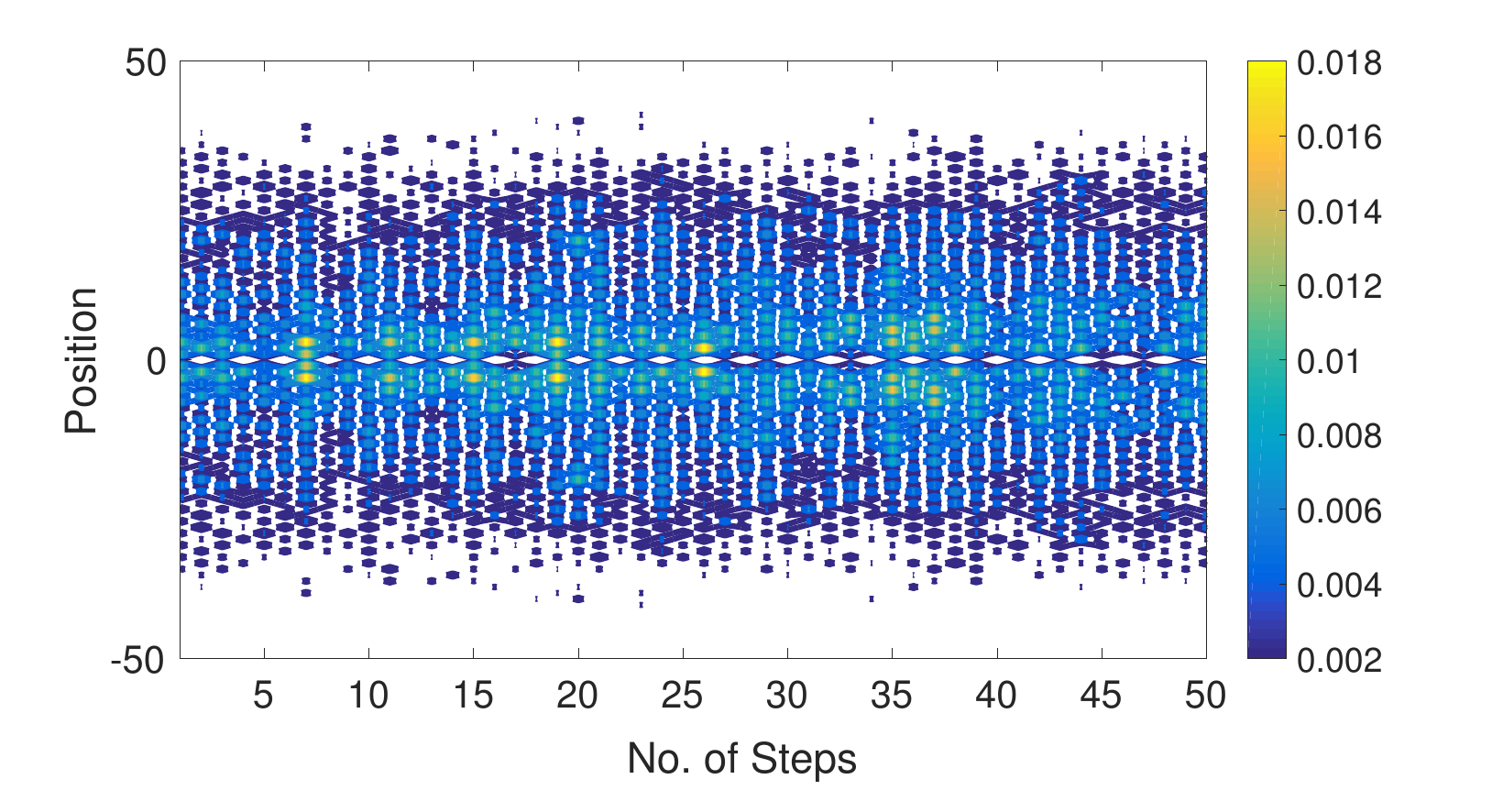} 
	\caption{Degree of interference ($\mu$) at each point in position space with respect to time-step in case of (a) HQW (b) SQW and (c) TQW, respectively, for $\textit{t} = 50$ and disordered system is averaged over 50-runs. The initial state is $\frac{1}{\sqrt{2}}(\ket{\uparrow}+\ket{\downarrow}) \otimes \ket{x=0}$.}
	\label{fig4}
\end{figure}

Therefore, the probability to find the walker at $x$ at time $(t+1)$ is given by,
\begin{align} \label{eq7}
P_{x,(t+1)} &= |\psi_{x,(t+1)}^{\uparrow}|^2  + |\psi_{x,(t+1)}^{\downarrow}|^2 \nonumber \\
&=\cos^2\theta \left(|\psi_{(x+1),t}^{\uparrow}|^2 + |\psi_{(x-1),t}^{\downarrow}|^2)\right) \nonumber \\
& + \sin^2\theta \left(|\psi_{(x-1),t}^{\uparrow}|^2 + |\psi_{(x+1),t}^{\downarrow}|^2\right) \nonumber \\
& + \sin\theta \cos\theta \left((\psi_{(x+1),t}^{\uparrow})^* \psi_{(x+1),t}^{\downarrow} \right. \nonumber \\
& \left. + (\psi_{(x+1),t}^{\downarrow})^* \psi_{(x+1),t}^{\uparrow}\right) \nonumber \\
& - \sin\theta \cos\theta \left( (\psi_{(x-1),t}^{\uparrow})^* \psi_{(x-1),t}^{\downarrow} \right. \nonumber \\
& \left. + (\psi_{(x-1),t}^{\downarrow})^* \psi_{(x-1),t}^{\uparrow}\right)
\end{align}
such that,
\begin{align}
P_{max} &= |\psi_{x,(t+1)}^{\uparrow}|^2  + |\psi_{x,(t+1)}^{\downarrow}|^2 \nonumber \\
&=\cos^2\theta \left(|\psi_{(x+1),t}^{\uparrow}|^2 + |\psi_{(x-1),t}^{\downarrow}|^2)\right) \nonumber \\
& + \sin^2\theta \left(|\psi_{(x-1),t}^{\uparrow}|^2 + |\psi_{(x+1),t}^{\downarrow}|^2\right) \nonumber \\
& + \sin\theta \cos\theta \left((\psi_{(x+1),t}^{\uparrow})^* \psi_{(x+1),t}^{\downarrow} \right. \nonumber \\
& \left. + (\psi_{(x+1),t}^{\downarrow})^* \psi_{(x+1),t}^{\uparrow}\right) \nonumber \\
& + \sin\theta \cos\theta \left( (\psi_{(x-1),t}^{\uparrow})^* \psi_{(x-1),t}^{\downarrow} \right. \nonumber \\
& \left. + (\psi_{(x-1),t}^{\downarrow})^* \psi_{(x-1),t}^{\uparrow}\right) \nonumber
\end{align}
and
\begin{align}
P_{min} &= |\psi_{x,(t+1)}^{\uparrow}|^2  + |\psi_{x,(t+1)}^{\downarrow}|^2 \nonumber \\
&=\cos^2\theta \left(|\psi_{(x+1),t}^{\uparrow}|^2 + |\psi_{(x-1),t}^{\downarrow}|^2)\right) \nonumber \\
& + \sin^2\theta \left(|\psi_{(x-1),t}^{\uparrow}|^2 + |\psi_{(x+1),t}^{\downarrow}|^2\right) \nonumber \\
& - \sin\theta \cos\theta \left((\psi_{(x+1),t}^{\uparrow})^* \psi_{(x+1),t}^{\downarrow} \right. \nonumber \\
& \left. + (\psi_{(x+1),t}^{\downarrow})^* \psi_{(x+1),t}^{\uparrow}\right) \nonumber \\
& - \sin\theta \cos\theta \left( (\psi_{(x-1),t}^{\uparrow})^* \psi_{(x-1),t}^{\downarrow} \right. \nonumber \\
& \left. + (\psi_{(x-1),t}^{\downarrow})^* \psi_{(x-1),t}^{\uparrow}\right) \nonumber
\end{align}
and therefore, $\mathcal{V}$ at position $x$ and at time $(t+1)$ is given by Eq.\,\eqref{eq8}.
\begin{widetext}

\begin{align} \label{eq8}
\mathcal{V}_{x,t+1} &\equiv \frac{P_{max} - P_{min}}{P_{max} + P_{min}} \nonumber \\
 &= \frac{\sin\theta \cos\theta \Big((\psi_{(x+1),t}^{\uparrow})^* \psi_{(x+1),t}^{\downarrow} + (\psi_{(x+1),t}^{\downarrow})^* \psi_{(x+1),t}^{\uparrow}+ (\psi_{(x-1),t}^{\uparrow})^* \psi_{(x-1),t}^{\downarrow} + (\psi_{(x-1),t}^{\downarrow})^* \psi_{(x-1),t}^{\uparrow}\Big)}{\cos^2\theta \left(|\psi_{(x+1),t}^{\uparrow}|^2 + |\psi_{(x-1),t}^{\downarrow}|^2)\right) + \sin^2\theta \left(|\psi_{(x-1),t}^{\uparrow}|^2 + |\psi_{(x+1),t}^{\downarrow}|^2\right)}.
\end{align}

\end{widetext}
The denominator in Eq.\,\eqref{eq8} corresponds to the sum of the probability densities at $(x+1)$ and $(x-1)$ at time $t$ due to the different coin states but the numerator are responsible for the interference visibility in the position space of the QW. Therefore, interference at point $x$ of position space at a given time-step $(t+1)$ can be visuals in terms of degree of interference ($\mu$) as,
\begin{align}\label{mu}
\mu_{x,t+1} &= \left|\sin\theta \cos\theta \left((\psi_{(x+1),t}^{\uparrow})^* \psi_{(x+1),t}^{\downarrow} \right. \right. \nonumber \\
&  + (\psi_{(x-1),t}^{\uparrow})^* \psi_{(x-1),t}^{\downarrow} + (\psi_{(x+1),t}^{\downarrow})^* \psi_{(x+1),t}^{\uparrow} \nonumber \\
& \left. \left. + (\psi_{(x-1),t}^{\downarrow})^* \psi_{(x-1),t}^{\uparrow}\right)\right|.
\end{align}
 where, $\psi^*$ is the conjugate of $\psi$ and for SQW and TQW, $\theta$ is position and time dependent, respectively. Fig.\,\ref{fig4} shows degree of interference ($\mu$) at each point in position space with respect to time (number of steps).
 
\section{\label{Sec4} Interference measure in localized QW}

QWs are known to interfere in position Hilbert space during the evolution and entangle the coin and position Hilbert space. Here we will try to quantify the role of interference in the localized states of QW. For this we calculate the amount of interference generated by different localized QW and identify the contribution of interference that leads to strongly localized states.    

Since for interference, system has to be coherent therefore Fig.\,\ref{fig2} shows the quantum coherence in complete Hilbert space $\mathcal{H} = \mathcal{H}_c \otimes \mathcal{H}_p$ for HQW and disorder induced localizations, SQW and TQW. When the coherence measure in Hilbert space $\mathcal{H}= \mathcal{H}_c \otimes \mathcal{H}_p$ is compared to the coherence measure in position and coin space separately as shown in Fig.\,\ref{fig3}, we see that the trend of complete coherence is mostly dominated by the coherence in the position space with some signature of coherence in coin space. The coherence in complete Hilbert space does not give any information of the coherence in coin Hilbert space. 

 Fig.\,\ref{fig3} right, shows the coherence in the coin space for the HQW, SQW and TQW, respectively, with time-step. Coherence in coin space is almost same for HQW and SQW. This implies that, walker's wave nature is same in both HQW and SQW. But in HQW, after the first few steps, the coherence in coin space saturates while in SQW, the coherence in coin space has the oscillatory behaviour with the time steps. This can be seen as an explanation for the spread and localization in HQW and SQW, respectively. This is same in topological localization as well as coherence in coin space can be seen in Fig.\,\ref{fig6} and corresponding probability distribution in Fig.\,\ref{fig7}. Stronger the localization, more is the oscillatory behaviour.    
 It is due to the fact that when the wave packet is localized around the origin, every alternative step, the walker prominently interfere constructively and destructively near the initial state. In TQW, we see a decrease in the coherence in coin space, which indicates that the walker's wave nature decreases with time step and is less than SQW and HQW. It is interesting to note that TQW is simulation of weak localization and SQW is simulation of strong (Anderson) localization and it is know from the previous studies that strong and weak localization is consequence of interference between the electron waves in the system. Stronger the localization more electron waves (wave nature of electron) are interfering. 

In position space, with increase in the number of steps, coherence increases for HQW, a small increase is seen for TQW and SQW with number of steps and very soon reaches a steady state value. For SQW, steady state is achieved faster than TQW and therefore coherence in position space is lowest for SQW. Now since from Eq.\,\eqref{particleNature}, it can be understand that the which-path information (particle nature) decreases with the increase in the coherence in the position space. Therefore, in HQW, walker has the lowest which-path information and hence the spread of the probability amplitude is maximum over the position space with time. The coherence in position space for SQW is minimum in time, therefore which-way information for SQW is more and hence the particle nature of walker in SQW is more compared to TQW or HQW and it should be obvious from the probability distribution as well since maximum localization is seen for SQW which is a simulation of strong localization. The coherence of TQW is minimum when compared to the coherence in position space of HQW with time, which implies walker in TQW is more particle in nature and hence more localized than HQW.  From this observation we can say that as the localization in QW increases, the coherence in position Hilbert space decreases. But the coherence in coin space shows alternative maxima and minima in localized QW while saturates to a constant value for HQW. Due to this, in localized case, interference happens near the point of localization but in homogeneous QW, it spreads over the position space, respectively. It is interesting to observe that duality relation in HQW, SQW and TQW as $\Big(D_Q + \dfrac{1}{N}C_c\Big)_{HQW}< \Big(D_Q + \dfrac{1}{N}C_c\Big)_{TQW} < \Big(D_Q + \dfrac{1}{N}C_c\Big)_{SQW} < 1$. 
The degree of interference ($\mu$) in Fig.\,\ref{fig4} for HQW, SQW and TQW, respectively, shows that for HQW interference happens all over the position space but intensity of interference measure is maximum at the extreme points with time-step representing that the probability distribution is spread all over the position space with probability peaks at the extreme points. For SQW, it shows that interference is happening near the initial state, implying probability amplitude is not moving away from the initial state and hence the walker is localized implying high visibility at the initial state.  Similarly for TQW, the degree of interference is all over the position space with maximum near the initial state. It is important to note that $\mu$ is the degree of interference that show the spread of interference in the system but not the interference visibility. 

 Similar explanation can be given for the quantization of interference in topological phases. Fig.\,\ref{fig7}-(a), \,\ref{fig7}-(c), \,\ref{fig7}-(e) and \,\ref{fig7}-(g) shows probability distribution for different topological phase and Fig.\,\ref{fig7}-(b), \,\ref{fig7}-(d), \,\ref{fig7}-(f) and \,\ref{fig7}-(h) shows corresponding coherence in complete Hilbert space $\mathcal{H} = \mathcal{H}_c \otimes \mathcal{H}_p$, respectively. Coherence in complete Hilbert space $\mathcal{H} = \mathcal{H}_c \otimes \mathcal{H}_p$ has the same trend as the coherence in position Hilbert space $\mathcal{H}_p$ as shown in Fig.\,\ref{fig6}-(a), \,\ref{fig6}-(c), \,\ref{fig6}-(e) and \,\ref{fig6}-(g) with the small signature of coherence of the coin Hilbert space $\mathcal{H}_c$ as shown in Fig.\,\ref{fig6}-(b), \,\ref{fig6}-(d), \,\ref{fig6}-(f) and \,\ref{fig6}-(h), respectively. Coherence in position space for topological localization also shows that as the localization in the system increases coherence in the position space decreases which implies more which-way information and hence walker's particle nature. Therefore, we can say that less the coherence in position space in an evolution process represents localization in the system and more the coherence in coin space shows the wave nature of the walker and more oscillatory behaviour in time shows the interference near the initial state which leads to localization. Fig.\,\ref{fig6}-(b), \,\ref{fig6}-(d),\,\ref{fig6}-(f) and \,\ref{fig6}-(h) shows coherence in coin space for different topological phases. It can be seen that for $(\theta_1, \theta_{2-}, \theta_{2+}) = (\pi/2, -3\pi/4, 3\pi/4)$ and $(\theta_1, \theta_{2-}, \theta_{2+}) = (-3\pi/2, -\pi, \pi)$, we get strongest localization as shown in Fig.\,\ref{fig7}-(c) and \,\ref{fig7}-(g), respectively and hence in this case we get the strongest and very clear alternate maxima and minima as shown in Fig.\,\ref{fig6}-(d) and \,\ref{fig6}-(h),respectively. An alternative constructive and destructive nature leads to the localization and minimum coherence in position space.    


\section{\label{Sec5} Conclusion}

QW which spreads quadratically faster in position space result in localization due to disorder in the evolution operator or due to topological effect. Interference during evolution plays a significant role both, for wide spread of wave packet in position space and for localization. By quantifying the interference, we showed that for localized states, the walker's particle nature is higher compared to the walker of homogeneous walk and in strong localization we get frequent alternative constructive and destructive interference and it is more clear from the coherence in the coin space which quantifies the wave nature of the walker. Major contribution comes from the coherence in the coin Hilbert space of QW, which quantifies the wave nature and hence the interference in the time. In topological QW, we get maximum localization at the initial position. In this case, the coherence in position space near the point of localization is observed to be zero, which in case gives maximum which-way information and hence the walker is mostly present at initial position and due to alternative maxima and minima of coherence in coin space the walker interferes in the neighbouring site of initial state.
 
We have also shown that the coherence in coin and position Hilbert space, quantifies the wave and particle nature of the walker, respectively, and can serve as an indicator of localization in position space and this will be very resourceful in localization studies. 
The coherence in coin Hilbert space shows maxima and minima at alternative step for localized system. This contributes to the constructive and destructive interference near the point of localization and which contributes to the decrease in the spread of probability distribution in the position Hilbert space.

\bibliographystyle{plain}

\end{document}